\documentclass[journal]{IEEEtran}

\usepackage{booktabs}
\usepackage{tabularx}
\usepackage{amssymb}
\usepackage{amsbsy}
\usepackage{amsmath}
\usepackage{bm}
\usepackage{verbatim}
\usepackage{mathrsfs}
\usepackage{amsfonts}
\usepackage{graphicx}
\usepackage[tight,footnotesize]{subfigure}
\usepackage[10pt]{moresize}
\usepackage{array}
\usepackage{color}
\usepackage{epsfig}
\usepackage{stfloats}
\usepackage{balance}

\usepackage{booktabs}
\usepackage{multirow}
\usepackage{makecell}
\usepackage{array}
\usepackage{cancel}

\usepackage[noend]{algpseudocode}
\usepackage{algorithmicx,algorithm}

\usepackage{setspace}
\usepackage{cases}

\usepackage{graphicx}
\usepackage{epstopdf}
\usepackage{multirow}
\usepackage{extarrows}

\usepackage{pifont} 
\usepackage{enumerate}
\usepackage{enumitem}

\newcommand{\subparagraph}{}
\usepackage{titlesec}
\usepackage[colorlinks]{hyperref}
\titlespacing{\section}{0pt}{2 ex plus .0ex minus .0ex}{1ex plus .0ex}
\titlespacing{\subsection}{0pt}{1.5 ex plus .0ex minus .0ex}{0.8 ex plus 0.0ex}
\titlespacing{\subsubsection}{0pt}{0.5ex plus .0ex minus .0ex}{0.0ex plus .0ex}

\usepackage{amsmath} 
\allowdisplaybreaks[4]

\ifCLASSINFOpdf

\else

\fi

\allowdisplaybreaks[4]

\begin{document}

\title{Split-Gate Pooled-Evidence Stochastic-Rollout Scheduling for Timely Progressive Edge Inference}

\author{Sai Xu,
Yinbo Yu,
Yanan~Du, 
Xusheng~Zhu, and
Gaojie~Chen
\vspace{-3mm} 

\thanks{S. Xu, and X. Zhu are with University College London, London, UK (e-mail: \texttt{sai.xu@ieee.org, xusheng.zhu@ucl.ac.uk}). Y.~Yu is with College of Artificial Intelligence, Nanjing University of Aeronautics and Astronautics, Nanjing, Jiangsu, 210016, China (e-mail: \texttt{yinboyu@nuaa.edu.cn}). Y.~Du is with the Department of Electronic and Electrical Engineering, University of Sheffield, Sheffield, S1 4ET, UK (e-mail: \texttt{yanan.du@ieee.org}). G. Chen is with the School of Flexible Electronics (SoFE), Sun Yat-sen University, Shenzhen, Guangdong 518107, China (e-mail: \texttt{gaojie.chen@ieee.org}).}
}

\maketitle

\begin{abstract}
This paper investigates causal radio scheduling for progressive edge inference, with the goal of maximizing timely inference throughput under job-specific deadlines. Specifically, a multi-tenant system is modeled in which each job alternates between wireless transmission and graphics processing unit (GPU) computation. The model captures time-varying uplink service, inter-stage precedence, variant-aware batching, non-preemptive execution on two GPU streams, and heterogeneous deadlines. To account for delayed radio--GPU coupling, a split-gate pooled-evidence stochastic-rollout (SGPE-SR) scheduler is proposed. Candidate and fallback policies are selected on one set of sampled futures, after which admission is evaluated on an independent held-out set; common random numbers are retained within each held-out candidate--fallback comparison. The override is executed only when its pooled nominal/recent-history held-out advantage is sufficiently positive and an independent recent-history replication statistic is nonnegative. System-level simulations using measured GPU profiles and paired random instances show that, in a prespecified long-horizon evaluation over 30 previously unused seeds and 3{,}379 offered jobs, SGPE-SR improves timely completions over the strongest rate-based baseline by 4.27\%, with a mean paired gain of 3.00 jobs per run and a 95\% bootstrap interval of $[1.60,\,4.43]$. 
\end{abstract}


\section{Introduction}\label{sec:introduction}

\IEEEPARstart{E}{dge} inference has emerged as a promising paradigm for supporting latency-sensitive perception services in cameras, robots, unmanned aerial vehicles, and industrial sensing systems~\cite{Letaief2022EdgeAI,Shao2022TaskOriented}. In these applications, resource-constrained end devices generate deep neural network (DNN) inference jobs and transmit the associated data over a shared wireless uplink to a graphics processing unit (GPU)-equipped edge server. The resulting end-to-end latency comprises wireless transmission, edge queueing, and GPU execution. Unlike conventional data delivery, an inference result is useful only if it is completed before its job-specific deadline. Therefore, maximizing \emph{timely inference throughput}, defined as the number of inference jobs completed within their respective deadlines, is a key objective in deadline-constrained edge intelligence~\cite{Shao2023Cooperative,Chen2024Realtime}.

Progressive inference, in which task-relevant information is transmitted and processed incrementally, provides a promising means of improving timely inference throughput while reducing communication and computation overhead~\cite{Shao2022TaskOriented,Shao2023Cooperative,Shao2024VideoAnalytics}. An end device first uploads a compact base representation for lightweight edge analysis. The resulting intermediate output is then used to identify an informative region of interest (ROI), which is subsequently transmitted for refined processing. This staged design follows the principle of communicating only task-relevant information rather than the complete input~\cite{Im2024AttentionAware}. By avoiding immediate transmission and processing of the full high-resolution input, progressive inference reduces redundant communication and computation. In addition, compatible inference workloads can be batched or executed concurrently to improve GPU efficiency.

Radio scheduling in such an alternating radio--GPU pipeline is inherently coupled with downstream computation. Serving a job at a radio stage determines when its next GPU stage becomes ready and thereby affects GPU queue composition, batching opportunities, concurrent execution, and waiting behind non-preemptive GPU waves. A radio decision may also alter the batching and deadline outcomes of other jobs. Consequently, the job with the highest instantaneous transmission rate is not necessarily the one whose service yields the largest timely inference throughput. These decisions must be made causally, without knowledge of future arrivals, job complexities, or channel realizations. Although prior work has investigated communication-efficient inference, heterogeneous offloading, and accelerator scheduling~\cite{Chen2024Realtime,Li2023Adaptive,Lin2023CoInference}, the specific setting studied here---candidate-wise causal radio evaluation through a two-upload/three-GPU-stage pipeline with fixed non-preemptive execution---has received less direct attention.

\subsection{Related Work}

Motivated by the need to capture the downstream effects of radio decisions in deadline-constrained progressive edge inference, this paper studies causal radio scheduling in an alternating radio--GPU pipeline. The most relevant studies fall into three categories: communication-efficient edge inference, GPU-aware inference scheduling, and prediction-assisted resource control.

\emph{1) Communication-efficient edge inference:} A substantial body of work reduces wireless overhead by transmitting compact task-oriented representations instead of raw inputs. Information-bottleneck-based approaches jointly optimize feature extraction and communication~\cite{Shao2022TaskOriented}, and have been extended to cooperative inference and edge video analytics~\cite{Shao2023Cooperative,Shao2024VideoAnalytics}. Other studies investigate progressive feature delivery~\cite{Lan2023Progressive}, selective transmission of informative visual regions~\cite{Im2024AttentionAware}, and joint optimization of model partitioning, offloading, and communication--computation resources~\cite{Chen2024Realtime,Li2024ModelSplitting,Lyu2024ObjectiveDriven}. These studies improve communication efficiency and device--edge collaboration, but typically assume a one-shot transfer, a single progressive transmission phase, or a fixed model partition. Consequently, they generally do not account for an alternating radio--GPU pipeline in which completion of a GPU stage may trigger a subsequent wireless transmission and alter the release times of downstream computation stages.

\emph{2) GPU-aware inference scheduling:} Shared inference systems improve accelerator utilization through adaptive batching, concurrency control, resource partitioning, and priority-aware scheduling. Representative systems such as VELTAIR, DVABatch, and Gpulet support heterogeneous inference workloads through compilation optimization, dynamic batching, and GPU sharing~\cite{Liu2022VELTAIR,Cui2022DVABatch,Choi2022Gpulet}, while REEF, TETRIS, and ElasticRoom provide mechanisms for preemption, memory sharing, and priority-aware execution~\cite{Han2022REEF,Li2022Tetris,Ma2024ElasticRoom}. More recent work further addresses unpredictable arrivals and multi-stage inference. SHEPHERD combines offline planning with online scheduling~\cite{Zhang2023Shepherd}, AlpaServe exploits model parallelism for statistical multiplexing~\cite{Li2023AlpaServe}, and vLLM enables efficient batching through dynamic management of inference states~\cite{Kwon2023vLLM}. Splitwise and DistServe separate inference phases to mitigate resource interference~\cite{Patel2024Splitwise,Zhong2024DistServe}, whereas Sarathi-Serve and Llumnix improve batching and load balancing through chunked execution and request migration~\cite{Agrawal2024Sarathi,Sun2024Llumnix}. These systems primarily optimize requests after they enter the accelerator subsystem and generally do not account for how upstream radio decisions shape GPU-stage release times, batching opportunities, and end-to-end deadline outcomes.

\emph{3) Prediction-assisted resource control:} Prediction- and model-assisted optimization has been widely used to anticipate variations in traffic, channel conditions, workloads, and service demand. Existing studies combine traffic prediction with resource allocation for split inference~\cite{Lyu2024ObjectiveDriven}, adapt uplink scheduling to predicted wireless conditions~\cite{Ren2024PredictiveUplink}, and enable proactive placement, offloading, and resource allocation based on estimated or digital-twin system states~\cite{Sfaxi2024ProactivePlacement,Chen2024Realtime,Wang2024DigitalTwin}. VIDUR further shows that predictive models can efficiently evaluate inference-serving configurations without executing every alternative on physical hardware~\cite{Agrawal2024Vidur}. These methods, however, mainly optimize aggregate resource allocation, service placement, or accelerator configurations. They generally do not evaluate candidate radio actions under shared sampled futures or explicitly propagate their downstream effects through GPU queues, compatible batching, concurrent non-preemptive execution, and multiple communication--computation stages.

In summary, prior studies have separately advanced task-oriented transmission, progressive feature delivery, GPU inference scheduling, and prediction-assisted resource allocation. Their joint treatment, however, remains limited in progressive edge inference, where a single radio decision may affect multiple downstream communication and GPU stages. This paper addresses this limitation by developing a causal scheduler that evaluates structured radio actions under shared conditional futures, propagates their consequences throughout the radio--GPU pipeline, and invokes rollout-based interventions only when they are supported by sufficient evidence relative to a baseline policy.

\subsection{Contributions}

To address the above cross-stage coupling, this paper proposes a split-gate pooled-evidence stochastic-rollout (SGPE-SR) scheduler for timely progressive edge inference. The main contributions are summarized as follows.

\begin{itemize}[leftmargin=*]

\item A multi-tenant progressive edge inference model is established in which each job alternates between two radio stages and three GPU stages. The model jointly captures time-varying uplink service, stage precedence, variant-aware batching, two-stream non-preemptive GPU execution, and job-specific deadlines. Based on this model, the radio-scheduling problem is formulated to maximize expected timely throughput over causal and feasible radio policies under a fixed GPU executor.

\item An SGPE-SR scheduler is developed to generate conditional futures from nominal and recent-history models using only causally available information. Candidate and fallback policies are selected using one future set, while admission is determined using an independent held-out set. This split preserves paired common-random-number comparisons without reusing the returns employed for policy selection as gate evidence. The selected rule replaces the fallback only when its pooled nominal/recent-history advantage is sufficiently positive and an independent recent-history replication statistic is nonnegative.

\item A paired system-level evaluation is conducted using 100-ms warm-up and 500-ms measurement windows, heterogeneous radio-rate and deadline classes, measured P95 GPU profiles, and five causal baselines. The evaluation separates a 30-seed design audit from a frozen 30-seed confirmation on previously unused seeds, reports paired-bootstrap confidence intervals and an exact sign-flip test, and examines load sensitivity over 220--300 jobs/s.

\end{itemize}

The remainder of this paper is organized as follows. Section~\ref{sec:system} presents the system model and problem formulation. Section~\ref{sec:method} develops SGPE-SR. Section~\ref{sec:evaluation} reports the system-level evaluation, and Section~\ref{sec:conclusion} concludes the paper.

\section{System Model and Problem Formulation}
\label{sec:system}

\subsection{System Overview}
\label{subsec:overview}

\begin{figure}[t]
    \centering
    \includegraphics[width=0.48\textwidth]{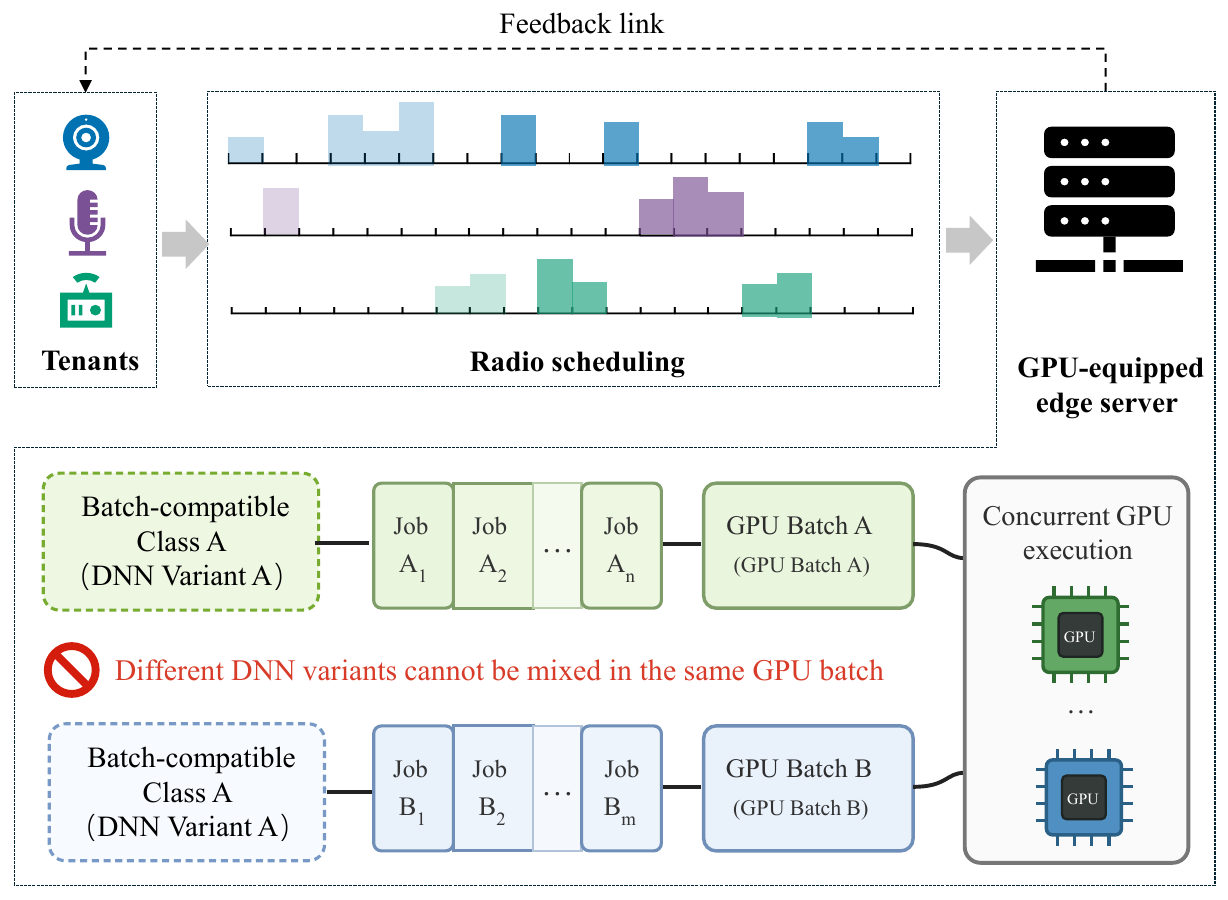}
    \caption{An illustration of the multi-tenant edge inference system.}
    \label{fig:system_model}
\end{figure}

Fig.~\ref{fig:system_model} illustrates a multi-tenant edge inference system comprising multiple tenants, each associated with an end device, and a GPU-equipped edge server. The end devices dynamically generate deadline-constrained DNN inference jobs and transmit the corresponding base and ROI data to the edge server over a shared wireless uplink. Each job alternates between wireless transmission and GPU computation stages. In each time slot, the radio scheduler serves at most one eligible job, while a fixed, non-preemptive GPU executor performs stage-dependent batching and concurrent execution. Jobs using different DNN variants belong to distinct batch-compatibility classes and therefore cannot be included in the same GPU batch. The model captures the coupling among time-varying wireless service, progressive stage releases, and downstream GPU execution. The objective is to maximize timely inference throughput, measured by the number of jobs completed within their respective deadlines, using only causally available information.

Let $\mathcal U=\{1,\ldots,U\}$, $\mathcal V=\{1,\ldots,V\}$, and $\mathcal J =\{1,\ldots,J\}$ denote the sets of tenants, DNN variants, and jobs, respectively. Time is slotted as $t\in\mathcal T=\{0,1,\ldots, T\}$ with slot duration $\Delta$. For job $j$, its attributes can be represented by
\begin{equation}
 J_j=\bigl(u_j,v_j,a_j,d_j,b_j^{\rm BU},b_j^{\rm RU},c_j\bigr),
 \label{eq:job_tuple}
\end{equation}
where $u_j\in\mathcal U$ and $v_j\in\mathcal V$ identify its tenant and DNN variant, respectively; $a_j$ is its arrival slot; and $d_j$ is its absolute deadline, by which the complete pipeline must finish. The payloads $b_j^{\rm BU}$ and $b_j^{\rm RU}$ specify the amounts of base and ROI data to be transmitted. The descriptor $c_j$ captures input-dependent content complexity and parameterizes the base and ROI payload sizes. The scheduler observes only jobs and attributes revealed by the current pipeline state; it
has no access to future job arrivals or their attributes.

\subsection{Progressive Inference}
\label{subsec:job_model}

Instead of offloading the entire input at once, progressive inference acquires the input through two transmissions, with an intermediate GPU result determining the content of the second. Specifically, each job $j\in\mathcal J$ traverses five stages. The end device
first transmits a compact input representation in the base-upload stage ($\mathrm{BU}$). The edge GPU then performs lightweight coarse analysis in the scout-GPU stage ($\mathrm{SG}$) to identify an informative ROI. Guided by the scout result, the end device transmits the selected ROI data in the ROI-upload stage ($\mathrm{RU}$). Finally, the GPU processes the ROI in the refinement-front stage ($\mathrm{RF}$) and produces the inference result in the refinement-head stage ($\mathrm{RH}$). Thus, communication and computation alternate within each job according to the following stage sequence:
\begin{equation}
\mathrm{BU}\prec\mathrm{SG}\prec\mathrm{RU}
\prec\mathrm{RF}\prec\mathrm{RH},
\label{eq}
\end{equation}
where $\prec$ denotes the precedence relation between consecutive stages.
The output of each stage is required by the next, and a job is complete only after its RH stage finishes.

Let $\mathcal K^{\rm R}=\{\mathrm{BU},\mathrm{RU}\}$ and $\mathcal K^{\rm G}=\{\mathrm{SG},\mathrm{RF},\mathrm{RH}\}$ denote the radio-transmission and GPU-computation stage sets, respectively, where the superscripts $\mathrm R$ and $\mathrm G$ stand for radio and GPU.
Because a job alternates between two resources, its stage must indicate both its pipeline progress and the resource it currently requires. Let $p_j(t)$ denote the state of job $j$ at the beginning of slot $t$:
\begin{equation}
p_j(t)\in
\mathcal K^{\rm R}\cup\mathcal K^{\rm G}
\cup\{\mathrm{done},\mathrm{drop}\}.
\label{eq:job_stage}
\end{equation}
If $p_j(t)\in\mathcal K^{\rm R}$, job $j$ is either waiting for uplink service or currently being served over the uplink. If $p_j(t)\in\mathcal K^{\rm G}$, it is either waiting for or executing its corresponding GPU stage. Upon successful completion of the $\mathrm{RH}$ stage, the job enters the absorbing state $\mathrm{done}$. The state $\mathrm{drop}$ indicates that the job has either missed its deadline or been discarded because on-time completion is no longer feasible.

Let $A_j^k$ and $C_j^k$ denote the release and completion slots of stage $k$, respectively. The base upload is released upon arrival, $A_j^{\rm BU}=a_j$, and every adjacent pair $(k,k^+)$ satisfies 
\begin{equation}
 A_j^{k^+}=C_j^k+1.
\label{eq:stage_release}
\end{equation}
Thus, a stage completed in slot $t$ releases its successor in slot $t+1$. The successor may start later because of radio or GPU contention, but such waiting occurs after release and does not change $A_j^{k^+}$. This precedence rule prevents stage skipping and ensures that one job cannot consume radio and GPU service simultaneously.

\subsection{Shared Uplink Transmission}
\label{subsec:radio_model}

The uplink is modeled as providing one orthogonal transmission opportunity per slot. This abstraction can represent a single channel, a resource block, or the output of a higher-layer scheduler after lower-layer resource allocation has been resolved. Let $R_u(t)$ denote the amount of data that tenant $u$ can transmit during slot $t$. This quantity is observed before the scheduling decision for slot $t$ and may vary across tenants and over time because of heterogeneous and time-varying wireless conditions.

For each job $j$, let $q_j(t)$ denote its remaining payload when it is in a radio stage. Define $e_j^{\rm tx}(t)\in\{0,1\}$ as the transmission-eligibility indicator, which equals one when job $j$ has arrived, remains active, is in a radio stage, and has a positive remaining payload, and equals zero otherwise. Let $x_j(t)\in\{0,1\}$ indicate whether job $j$ is scheduled for uplink transmission in slot $t$. The feasible uplink action set is
\begin{equation}
\begin{aligned}
\mathcal X(t)=\Bigl\{\boldsymbol{x}(t)\in\{0,1\}^{|\mathcal J|}:\
&\sum_{j\in\mathcal J}x_j(t)\le 1,\\
&x_j(t)\le e_j^{\rm tx}(t),\quad \forall j\in\mathcal J
\Bigr\}.
\end{aligned}
\label{eq:radio_feasibility}
\end{equation}
Thus, at most one eligible radio-stage job can be served in each slot, while the uplink may remain idle when no such job is available.
If job $j$ is scheduled in slot $t$, its remaining payload evolves as
\begin{equation}
q_j(t+1)
=
\bigl[q_j(t)-x_j(t)R_{u_j}(t)\bigr]^+,
\quad
[z]^+\triangleq\max\{z,0\}.
\label{eq:payload_update}
\end{equation}
The radio stage completes when $q_j(t+1)$ first reaches zero, after which its GPU successor is released according to \eqref{eq:stage_release}. Radio service is preemptive at slot boundaries: an unfinished job retains its residual payload, and any eligible job may be scheduled in the next slot. Hence, the radio scheduling sequence constitutes the primary control variable.

\subsection{Fixed Edge GPU Execution}
\label{subsec:gpu_model}

Completion of a radio stage releases the corresponding inference workload to the edge GPU, whose queueing and execution determine the remaining latency. For each GPU stage $g\in\mathcal K^{\rm G}$ and DNN variant $v\in\mathcal V$, let $\mathcal Q_{g,v}(t)$ denote the set of ready jobs waiting to execute stage $g$. Jobs are partitioned by $(g,v)$ because only those executing the same stage of the same DNN variant may be batched together. Accordingly, a feasible batch satisfies
\begin{equation}
B\subseteq\mathcal Q_{g,v}(t),
\qquad
1\le |B|\le B_{\max},
\label{eq:feasible_batch}
\end{equation}
where $B_{\max}$ is the maximum supported batch size. Batching improves GPU efficiency but introduces a latency--throughput tradeoff: launching a small batch reduces waiting time, whereas delaying execution to collect more compatible jobs may improve throughput at the expense of deadline slack.

The GPU provides two independent execution streams. A \emph{wave} denotes the set of batches launched simultaneously whenever the GPU becomes idle. Let $\mathcal B_t$ denote the set of feasible batches at slot $t$. A legal wave consists of either one batch or two disjoint, concurrency-compatible batches:
\begin{equation}
\begin{aligned}
\mathcal W_t=\bigl\{&(B_1),(B_1,B_2):
B_1,B_2\in\mathcal B_t,\\
&B_1\cap B_2=\varnothing,\ 
\chi(B_1,B_2)=1
\bigr\},
\end{aligned}
\label{eq:feasible_wave}
\end{equation}
where $\chi(B_1,B_2)=1$ indicates that the fixed GPU configuration permits concurrent execution. The duration of a single batch $B$ is given by the offline profile $P(g,v,|B|)$, while the stream-specific durations of a concurrent pair are given by $P^{\rm con}(B_1,B_2)$. Both profiles are measured offline and treated as fixed system parameters, rather than being estimated or optimized online.

Let $G_t$ collect the current GPU state, including the running wave, constituent batches, residual stream durations, ready queues, and ready-job ages. The implemented executor has no batching-wait timer: it is invoked whenever the GPU is idle and at least one legal batch is ready. For each GPU stage and compatibility label, ready jobs are sorted by $(d_j,j)$, and the earliest-deadline prefix of every profiled size in $\{1,2,4\}$ is considered. Feasible one-batch waves and all disjoint two-batch waves with a measured concurrency profile are enumerated. A wave is discarded if either its measured completion time or an optimistic bound on the remaining radio and GPU work makes any constituent job late. Among the remaining waves, the executor lexicographically minimizes
\begin{equation}
\left(
s_{\min}(W),\
\frac{T(W)}{|J(W)|},\
-|J(W)|,\
\mathbf 1\{|W|=1\},\
\kappa(W)
\right),
\label{eq:gpu_wave_key}
\end{equation}
where $s_{\min}(W)$ is the minimum optimistic post-wave slack, $T(W)$ is the measured wave makespan, $J(W)$ is its job set, and $\kappa(W)$ is the deterministic tuple of stage, compatibility label, job indices, and stream durations. Hence, deadline urgency is primary, followed by per-job makespan, larger waves, two-stream execution, and deterministic structural tie breaking. When a wave is running, no new batch is launched until both streams in that wave have completed. The resulting executor is
\begin{equation}
\Gamma_t=\pi_{\rm G}(G_t,\mathcal R_t),
\label{eq:fixed_gpu_policy}
\end{equation}
where $\mathcal R_t$ is the released GPU-stage set and $\Gamma_t$ is the launch or continuation decision. The executor and all measured profiles are fixed across radio policies. This setup isolates radio scheduling and makes GPU execution deterministic once GPU-stage release times are specified; the optimization therefore does not include joint radio--GPU control.
 
\subsection{Causal System Dynamics}
\label{subsec:causal_dynamics}

The job, radio, and GPU models are integrated into a unified system state. Immediately before the radio scheduling decision in slot $t$, the state is defined as
\begin{equation}
S_t=\bigl(S_t^{\rm job},S_t^{\rm R},S_t^{\rm G}\bigr),
\label{eq:system_state}
\end{equation}
where $S_t^{\rm job}$ records, for each revealed job, its current stage, remaining payload, release time, deadline, and terminal status; $S_t^{\rm R}$ contains the current uplink state and its observation history; and $S_t^{\rm G}=G_t$ represents the GPU state. The state spans both resources because radio-side information alone is insufficient to determine whether an uploaded job will immediately join a GPU batch or wait behind a running wave.

Let $\omega_{t+1}$ collect the exogenous events occurring between slots $t$ and $t+1$, including new job arrivals, newly revealed job attributes, and the next channel state. Given the radio action
$\boldsymbol{x}(t)=\{x_j(t):j\in\mathcal J\}$ and the fixed GPU executor, the system evolves according to
\begin{equation}
S_{t+1}
=
F\bigl(
S_t,\boldsymbol{x}(t),
\pi_{\rm G}(G_t,\mathcal R_t),
\omega_{t+1}
\bigr).
\label{eq:system_transition}
\end{equation}
The transition function $F$ is not an additional decision rule; it compactly captures the previously defined payload evolution, stage releases, GPU queueing, batch execution, and deadline-based termination. Therefore, \eqref{eq:system_transition} makes explicit that radio decisions influence future outcomes only through the physical processing pipeline and subsequent exogenous events.

Online decisions must depend only on information available at the decision time. Let $\mathcal F_t$ denote the information available immediately before slot $t$, including all jobs and attributes revealed by that time, the observed channel history, the running GPU wave and residual stream durations, the ready-job sets, and the fixed system parameters. It excludes future job arrivals, unrevealed attributes, and future channel realizations. An admissible radio policy must therefore satisfy
\begin{equation}
\boldsymbol{x}(t)=\pi_t(\mathcal F_t),
\qquad \forall t.
\label{eq:causality}
\end{equation}
Let $\Pi_{\rm causal}$ denote the set of policies satisfying \eqref{eq:causality} and the radio feasibility constraints. A model-based scheduler may generate conditional future samples based on $\mathcal F_t$, but it cannot access the unrevealed portion of the realized evaluation trace. This distinction separates causal prediction from noncausal offline planning.

\subsection{Problem Formulation}
\label{subsec:problem}

The service objective is defined in terms of complete end-to-end inference results rather than transmitted data volume or completed intermediate stages. Let $f_j^\pi$ denote the slot in which job $j$ completes its final $\mathrm{RH}$ stage under policy $\pi$. The corresponding utility is
\begin{equation}
y_j^\pi=\mathbf 1\{f_j^\pi\le d_j\},
\label{eq:timely_indicator}
\end{equation}
where $\mathbf 1\{\cdot\}$ denotes the indicator function. A job contributes one if it completes by its deadline and zero otherwise, including when it is completed late or dropped. No additional utility is assigned to earlier completion, reflecting applications with hard timeliness requirements.
For an arrival-accounting horizon $\mathcal H$, define $\mathcal J_{\mathcal H}=\{j:a_j\in\mathcal H\}$ as the set of jobs arriving within $\mathcal H$. Each such job is evaluated against its own deadline, even if its completion or expiration occurs after the end of $\mathcal H$. Thus, the horizon determines which arrivals are counted but does not impose an artificial common deadline. The timely throughput under policy $\pi$ is
\begin{equation}
Y^\pi(\mathcal H)
=
\sum_{j\in\mathcal J_{\mathcal H}}
\mathbf 1\{f_j^\pi\le d_j\}.
\label{eq:timely_throughput}
\end{equation}

Let $\mathbb P$ denote the generally unknown joint distribution of future arrivals, job attributes, and channel evolution. Although only the radio scheduler is optimized, radio decisions affect downstream GPU execution through stage releases. Consequently, maximizing instantaneous uplink service does not necessarily maximize timely inference throughput. Therefore, the radio-scheduling problem is formulated to select a causal policy that maximizes expected timely throughput, subject to the progressive processing pipeline and the fixed GPU executor:
\begin{subequations}
\label{prob:timely_scheduling}
\begin{align}
(\text{P1})~\max_{\pi\in\Pi_{\rm causal}}\quad  
&\mathbb E_{\mathbb P}\!\left[
\sum_{j\in\mathcal J_{\mathcal H}}
\mathbf 1\{f_j^\pi\le d_j\}
\right]
\label{prob:objective}\\
\operatorname{s. t.}\quad
&\eqref{eq:stage_release},\
\eqref{eq:radio_feasibility},\
\eqref{eq:payload_update},
\label{prob:radio_pipeline}\\
&\eqref{eq:feasible_batch},\
\eqref{eq:feasible_wave},\
\eqref{eq:fixed_gpu_policy},
\label{prob:gpu_constraints}\\
&\eqref{eq:system_transition},\
\eqref{eq:causality}.
\label{prob:dynamics_causality}
\end{align}
\end{subequations}
The constraints in \eqref{prob:radio_pipeline} govern stage precedence, uplink feasibility, and payload evolution. Those in \eqref{prob:gpu_constraints} specify feasible batching, concurrent execution, and the fixed non-preemptive GPU behavior. Finally, \eqref{prob:dynamics_causality} propagates radio actions through the complete radio--GPU pipeline and restricts each decision to causally available information. Consequently, the only optimized control is the radio-scheduling policy.

\section{Scheduling Method}
\label{sec:method}

The problem~(\text{P1}) is difficult because a radio action can change several later GPU releases, batch formations, and non-preemptive
waiting intervals, while the arrivals and channels that determine those effects are unavailable at decision time. Exhaustive optimization over
slot-level action sequences is combinatorial. To address the problem~(\text{P1}), this section develops SGPE-SR, which evaluates five structured radio rules over sampled end-to-end trajectories. At each planning epoch, it records the current physical state, causal context, and virtual deficits, then constructs independent selection and held-out gate sets from nominal and recent-history models. The selection set chooses both a macro rule and a Max-Rate/DPP (described below) fallback. The held-out set subsequently determines whether the frozen rule may override the frozen fallback. Otherwise, the causal fallback acts. The method is a causal model-based approximation and does not guarantee global optimality or realized-future safety.

\subsection{Causal Fallback Policies}
\label{subsec:fallback_policies}

Two deterministic causal policies are available as fallback branches. \emph{Max-Rate} selects the feasible job whose tenant has the largest
currently observed rate, with deadline, arrival time, and job index used for deterministic tie breaking:
\begin{equation}
j_{\rm MR}(t)
=\underset{j:e_j^{\rm tx}(t)=1}{\arg\min}
\bigl(-R_{u_j}(t),d_j,a_j,j\bigr).
\label{eq:max_rate_fallback}
\end{equation}
Max-Rate is the default because it requires no workload prediction and represents conventional channel-driven radio scheduling.
The second policy is a fixed virtual-deficit heuristic inspired by the drift-plus-penalty (DPP) principle. Let $Z_u(t)\ge0$ be tenant $u$'s service
deficit, and let $N_u^{\rm arr}(t)$ and $N_u^{\rm on}(t)$ be its newly arrived and newly completed-on-time jobs in slot $t$. The deficit evolves as
\begin{equation}
 Z_u(t+1)=
 \left[Z_u(t)+\rho N_u^{\rm arr}(t)-N_u^{\rm on}(t)\right]^+,
 \label{eq:deficit_update}
\end{equation}
where $\rho\in[0,1]$ is a target timely-completion fraction and $Z_u(0)=0$. Only realized arrivals and completions update the real deficit; simulated branches use private copies. This update is used as a scheduling heuristic: no deficit-stability claim is made, and the target need not be feasible when the offered workload exceeds system capacity.

For an eligible radio-stage job $j$, define
\begin{equation}
 \widehat n_j^{\rm cur}(t)=
 \begin{cases}
 \left\lceil q_j(t)/R_{u_j}(t)\right\rceil,
     &R_{u_j}(t)>0,\\
 +\infty, &R_{u_j}(t)=0,
 \end{cases}
 \label{eq:current_upload_slots}
\end{equation}
as the optimistic number of slots required to finish its current upload. The
optimistic remaining radio time is
\begin{equation}
 \widehat n_j^{\rm rem}(t)=
 \widehat n_j^{\rm cur}(t)
 +\mathbf 1\{p_j(t)=\mathrm{BU}\}
 \left\lceil\frac{b_j^{\rm RU}}{R_{u_j}(t)}\right\rceil,
 \label{eq:remaining_radio_slots}
\end{equation}
where the second term accounts for the later ROI upload when the job is still
in BU. Let $\widehat g_j^{\rm rem}(t)$ be the sum of the fastest legal
profile times of its remaining GPU stages. The projected slack is
\begin{equation}
 \widehat s_j(t)=
 d_j-t-\widehat n_j^{\rm rem}(t)-\widehat g_j^{\rm rem}(t).
 \label{eq:projected_slack}
\end{equation}
Jobs with $\widehat s_j(t)<0$ are removed by the common optimistic feasibility filter. The DPP heuristic selects the remaining job maximizing
\begin{equation}
 \phi_j^{\rm DPP}(t)=
 Z_{u_j}(t)
 \min\!\left\{1,\frac{R_{u_j}(t)}{q_j(t)}\right\}
 +\frac{\beta}{1+\widehat s_j(t)},
 \label{eq:dpp_score}
\end{equation}
with deterministic tie breaking. This expression is a fixed heuristic for the complete multi-stage system; no exact Lyapunov-optimality claim is made.

The stochastic rollout model also selects the fallback branch supported by the selection-sample statistic. Under each sampled selection future, both fallback branches are evaluated. Let $D_{B,r,n}=R_{{\rm DPP},r,n}-R_{{\rm MR},r,n}$ and let $L_{B,r}$ be its one-sided empirical lower statistic, defined as in
\eqref{eq:empirical_lower_statistic}. The selected fallback is
\begin{equation}
B_t=
\begin{cases}
{\rm DPP}, & L_{B,r}>0,\ \forall r\in\{\mathrm N,\mathrm R_1,\mathrm R_2\},\\
{\rm MR}, & \text{otherwise}.
\end{cases}
\label{eq:fallback_selection}
\end{equation}
Thus, Max-Rate remains the conservative default; DPP is used only when its simulated advantage is positive in every selection-future regime. The chosen fallback and all tie-breaking rules are frozen before held-out gate evaluation.

\subsection{Causal Conditional Future Model}
\label{subsec:conditional_model}

SGPE-SR constructs future scenarios using only the information available in $\mathcal F_t$ as defined in \eqref{eq:causality}. Specifically, its rolling context contains the most recent $H_{\rm ctx}$ revealed slots, including arrival counts, per-tenant rates, arrival dispersion, aggregate rate variations, zero-rate frequency, and the observed multi-tenant rate vectors. To preserve causality, the context excludes trace identifiers, absolute trace timestamps, dataset-partition labels, and all unrevealed arrivals, job
attributes, and channel states. Based on this revealed context, SGPE-SR uses a nominal model and two independent sample partitions from the same recent-history model:
\begin{equation}
\widehat{\mathbb P}_r(\cdot\mid\mathcal F_t),
\qquad
r\in\{\mathrm N,\mathrm R_1,\mathrm R_2\},
\label{eq:future_regimes}
\end{equation}
where $\mathrm N$ is the \emph{nominal} regime and $\mathrm R_1,\mathrm R_2$ are disjoint partitions sampled from the same \emph{recent-history} conditional model.

The nominal regime is a frozen reference model whose arrival intensity, complexity distribution, compatibility-label distribution, and AR(1) channel parameters are fixed using development experiments before the evaluation seeds are accessed. It is initialized only from the currently revealed state. Future arrivals follow the nominal Poisson model, job complexity and labels follow their fixed reference distributions, and rates follow the frozen AR(1) model. The recent-history model adapts this reference using only revealed context: its arrival intensity combines the nominal and empirical intensities, while complete observed multi-tenant rate vectors are resampled to retain recent cross-tenant dependence. The two recent-history partitions are independent conditional draws used to avoid relying on a single adaptive sample partition. All partitions begin from the same current jobs and GPU state and use the same fixed executor $\pi_{\rm G}$ and hardware profiles.

The regime-specific arrival intensities are defined as follows. Let $\lambda_0$ denote the frozen nominal arrival intensity. Using the latest $H_t\le H_{\rm ctx}$ revealed slots, the empirical arrival intensity is 
\begin{equation}
\lambda_{\rm obs}(t)
=
\frac{1}{H_t\Delta}
\sum_{\tau=t-H_t+1}^{t}
\sum_{u\in\mathcal U}N_u^{\rm arr}(\tau).
\label{eq:observed_arrival_rate}
\end{equation}
The recent-history intensity averages the nominal and empirical intensities:
\begin{equation}
\lambda_{\rm R}(t)
=
\max\!\left\{
\lambda_{\min},\frac{\lambda_0+\lambda_{\rm obs}(t)}{2}
\right\}.
\label{eq:recent_arrival_rate}
\end{equation}
All intensities are measured in jobs/s; $\lambda_{\min}=1$ job/s prevents a degenerate zero-arrival rollout without introducing a unitless lower bound.
Each sampled future is evaluated over a finite set of decision-relevant jobs. Let $\mathcal J_t^{\rm act}$ denote the unresolved accounted jobs revealed by time $t$. For regime $r$ and sample $n$, the target cohort additionally includes jobs sampled to arrive within the next $W$ slots:
\begin{equation}
\mathcal C_t^{(r,n)}
=
\mathcal J_t^{\rm act}
\cup
\left\{
j:
t<a_j^{(r,n)}
\le
\min\{t+W,T_{\mathcal H}\}
\right\},
\label{eq:target_cohort}
\end{equation}
where $T_{\mathcal H}$ is the end of the arrival-accounting horizon. Let $d_{\max}^{\rm rel}$ be the largest configured relative deadline. Future arrivals are generated only through $T_{\rm gen}=\min\{T_{\mathcal H}-1,t+W+d_{\max}^{\rm rel}\}$; no job is added after $T_{\rm gen}$. Noncohort jobs arriving before the branch stops may consume radio or GPU resources, but their outcomes do not enter the return. A cohort job becomes \textsc{done} when its RH stage completes by its deadline and becomes \textsc{drop} at its deadline otherwise. Every branch stops when all cohort jobs are terminal, no later than $T_{\rm stop}=\max_{j\in\mathcal C_t^{(r,n)}}d_j$. These generation and stopping rules match the implementation and guarantee a finite rollout while evaluating complete cohort deadline outcomes.
 
\subsection{Structured Radio Macro Rules}
\label{subsec:macro_rules}

Enumerating all job-level scheduling sequences over multiple slots leads to exponential growth in the action space. To keep planning finite, SGPE-SR evaluates a small set of five deterministic radio priority rules, indexed by
\begin{equation}
\mathcal A=\{0,1,2,3,4\}.
\label{eq:macro_action_set}
\end{equation}
Each element $a\in\mathcal A$ identifies a scheduling rule rather than a specific job. Once selected, the rule is applied for $L$ consecutive slots. At every slot, however, it re-ranks the currently eligible jobs using the latest causally available system state.
In addition to $\widehat n_j^{\rm cur}(t)$ and $\widehat s_j(t)$, let $m_j(t)$ denote the number of GPU-ready jobs that would be batch-compatible with job $j$ after its current upload is completed. Define
\begin{equation}
\eta_j(t)=\frac{R_{u_j}(t)}{q_j(t)}
\end{equation}
as the current radio efficiency of job $j$. Each rule selects, among the eligible jobs that pass the optimistic feasibility filter, the job with the lexicographically smallest priority key listed in Table~\ref{tab:macro_rules}.

\begin{table}[t]
\centering
\footnotesize
\caption{Structured Radio Macro Rules}
\label{tab:macro_rules}
\begin{tabular}{c l l}
\toprule
$a$ & Scheduling objective & Lexicographic priority key\\
\midrule
0 & Minimize projected slack
  & $(\widehat s_j,-R_{u_j},j)$\\
1 & Complete the current upload
  & $(\widehat n_j^{\rm cur},\widehat s_j,-R_{u_j},d_j,j)$\\
2 & Promote GPU batch formation
  & $(-m_j,\widehat n_j^{\rm cur},\widehat s_j,-R_{u_j},j)$\\
3 & Release an urgent GPU-stage job
  & $(\widehat s_j,\widehat n_j^{\rm cur},-m_j,-R_{u_j},j)$\\
4 & Exploit radio efficiency
  & $(-\eta_j,\widehat s_j,\widehat n_j^{\rm cur},j)$\\
\bottomrule
\end{tabular}
\end{table}

Rule~0 prioritizes jobs with the smallest projected end-to-end slack and serves as the conservative reference within the candidate rule set. Rule~1 favors jobs whose current uploads can be completed quickly. Rule~2 prioritizes jobs whose completion is likely to enlarge a compatible GPU-ready set and thereby facilitate batching. Rule~3 gives primary priority to deadline urgency while using upload time and batching potential as secondary criteria. Rule~4 favors jobs that can make the greatest relative transmission progress under the current uplink condition. Rule~0 serves as the common rollout suffix and as the within-set reference candidate; it is distinct from Max-Rate and DPP, which form the causal fallback set.
 
\subsection{Paired Counterfactual Evaluation}
\label{subsec:counterfactual_eval}

SGPE-SR evaluates all candidate and fallback actions under the same sampled exogenous future. In each scenario, the five macro-rule branches,
Max-Rate branch, and DPP branch share the same arrivals, attributes, channel trajectory, initial state, and fixed GPU executor. This common-random-number
construction isolates action-dependent differences from sampling noise.

For regime $r$, scenario $n$, and macro rule $a$, the candidate branch starts from a clone of the current system state. It applies rule $a$ for the next $L$ slots and then switches to Rule~0 as a common suffix until every job in the target cohort is resolved. The fixed GPU executor is used throughout the rollout. The resulting return is the number of cohort jobs completed by their deadlines:
\begin{equation}
R_{a,r,n}
=
\sum_{j\in\mathcal C_t^{(r,n)}}
\mathbf 1\!\left\{f_{j,a,r,n}\le d_j\right\}.
\label{eq:candidate_return}
\end{equation}
Two reference branches start from identical environment clones. The Max-Rate branch and the DPP branch apply their respective policies for
the first $L$ slots and then use the same Rule~0 suffix. Their returns are $R_{{\rm MR},r,n}$ and $R_{{\rm DPP},r,n}$. After selecting $B_t$ by
\eqref{eq:fallback_selection}, define $R_{B,r,n}=R_{B_t,r,n}$. The paired comparison therefore estimates the complete target-cohort consequence of replacing the next $L$ fallback decisions with one macro rule; it is not a comparison of policies executed permanently in the simulated suffix.

Candidate selection uses only the nominal futures in a selection set $\Omega_t^{\rm sel}$. Let $\mathcal I_{\rm N}^{\rm sel}$ denote its nominal index set, and let $\mathcal I_{\rm N}^{\rm sel}(N)$ denote the nominal indices within a total selection prefix of size $N$. For the selected prefix, define
\begin{equation}
\widehat\mu_a
=
\frac{1}{|\mathcal I_{\rm N}^{\rm sel}|}
\sum_{n\in\mathcal I_{\rm N}^{\rm sel}}
R_{a,\mathrm N,n}.
\label{eq:nominal_mean_return}
\end{equation}
Let $a^\star$ be the deterministically tie-broken maximizer of \eqref{eq:nominal_mean_return}. To avoid replacing the candidate-set reference for a negligible simulated gain, the rollout candidate is selected as
\begin{equation}
a_W
=
\begin{cases}
0,
& \displaystyle
\max_{a\in\mathcal A}\widehat\mu_a
\le \widehat\mu_0+\delta,\\[1mm]
a^\star,
& \text{otherwise},
\end{cases}
\label{eq:rollout_candidate}
\end{equation}
where $\delta$ is a fixed indifference margin. \eqref{eq:rollout_candidate} determines only the candidate rule. Whether that candidate may replace the selected Max-Rate/DPP fallback is decided by the subsequent evidence gate.

\subsection{\texorpdfstring{Independent Held-Out Evidence and Admission}{Independent Held-Out Evidence and Admission}}
\label{subsec:risk_gate}

At each planning epoch, the controller draws two mutually independent scenario sets conditional on the same causal information:
\begin{equation}
\Omega_t^{\rm sel}\perp\Omega_t^{\rm gate}
\mid\mathcal F_t.
\label{eq:independent_scenario_sets}
\end{equation}
The candidate rule and Max-Rate/DPP fallback are selected exclusively from $\Omega_t^{\rm sel}$. Both choices are then frozen before any return in $\Omega_t^{\rm gate}$ is used. Common random numbers are retained within each held-out scenario, so the frozen candidate and frozen fallback experience the same arrivals, attributes, channels, initial state, and GPU executor. For held-out regime $r$ and scenario $n$, define
\begin{equation}
D_{r,n}^{\rm gate}
=
R_{a_W,r,n}^{\rm gate}-R_{B_t,r,n}^{\rm gate}.
\label{eq:paired_advantage}
\end{equation}
A positive value means that the frozen candidate completes more target-cohort jobs on time than the frozen fallback under the same held-out future. Given $N_r^{\rm gate}$ held-out samples in regime $r$, define
\begin{equation}
\overline D_r^{\rm gate}
=
\frac{1}{N_r^{\rm gate}}\sum_{n=1}^{N_r^{\rm gate}}D_{r,n}^{\rm gate},
\qquad
{\rm se}_r^{\rm gate}
=
\frac{s(D_r^{\rm gate})}{\sqrt{N_r^{\rm gate}}},
\label{eq:advantage_statistics}
\end{equation}
and the empirical one-sided held-out statistic
\begin{equation}
L_r^{\rm gate}
=
\overline D_r^{\rm gate}-z_{\rm one}{\rm se}_r^{\rm gate},
\qquad
 r\in\{\mathrm N,\mathrm R_1,\mathrm R_2\},
\label{eq:empirical_lower_statistic}
\end{equation}
where $z_{\rm one}>0$ is fixed before evaluation. Sample splitting prevents the selected winner from being admitted on the same returns that selected it. Nevertheless, $L_r^{\rm gate}$ remains a model-based admission statistic, not a distribution-free confidence bound or a guarantee for realized-environment performance.

Within $\Omega_t^{\rm sel}$, a nested prefix of size $N_0^{\rm sel}$ and full set of size $N_1^{\rm sel}$, with $N_0^{\rm sel}<N_1^{\rm sel}$, are used only to assess selection stability. They produce $(a_{N_0},B_{N_0})$ and $(a_{N_1},B_{N_1})$. Let $L_{B,r}^{(N_0)}$ and ${\rm se}_{B,r}^{(N_0)}$ be the selection-prefix statistic and standard error for DPP minus Max-Rate. The selected depth is
\begin{equation}
\hspace{-1em}
N_{\rm sel}
=
\begin{cases}
N_1^{\rm sel},
& a_{N_0}\ne a_{N_1}\ \text{or}\ B_{N_0}\ne B_{N_1},\\
N_1^{\rm sel},
& \substack{L_{B,r}^{(N_0)}\text{ is nonfinite, or}\\
\left|L_{B,r}^{(N_0)}\right|
\le z_{\rm one}{\rm se}_{B,r}^{(N_0)}
\text{ for some }r},\\
N_0^{\rm sel},
& \text{otherwise},
\end{cases}
\label{eq:effective_evidence_size}
\end{equation}
so the full selection set is used when a discrete choice changes or the prefix cannot distinguish the fallbacks. The frozen choices are
\begin{equation}
(a_W,B_t)
=
\begin{cases}
(a_{N_1},B_{N_1}), & N_{\rm sel}=N_1^{\rm sel},\\
(a_{N_0},B_{N_0}), & N_{\rm sel}=N_0^{\rm sel}.
\end{cases}
\label{eq:final_nested_candidate}
\end{equation}
The implementation uses total selection-prefix and full-set sizes $N_0^{\rm sel}=32$ and $N_1^{\rm sel}=64$, together with an independent held-out set of total size $N^{\rm gate}=64$, preserving the original total of 128 materialized futures per planning decision. The held-out set is partitioned into $N_{\rm N}^{\rm gate}=21$, $N_{\rm R_1}^{\rm gate}=22$, and $N_{\rm R_2}^{\rm gate}=21$ futures, so that $\sum_r N_r^{\rm gate}=N^{\rm gate}=64$; the three partitions are not separate 64-future sets. A shorter selected depth changes only which selection returns determine $(a_W,B_t)$; all 64 held-out futures are always reserved for admission.

SGPE-SR pools nominal evidence with the first recent-history partition, while the second recent-history partition provides an independent replication check. The pooled statistic is
\begin{equation}
L_{\rm pool}^{\rm gate}
=
\frac{L_{\rm N}^{\rm gate}+L_{\rm R_1}^{\rm gate}}{2}.
\label{eq:pooled_evidence}
\end{equation}
The candidate is admitted only if all required statistics are finite and
\begin{equation}
L_{\rm pool}^{\rm gate}>m_{\rm gain},
\qquad
L_{\rm R_2}^{\rm gate}\ge 0,
\label{eq:rbpe_gate}
\end{equation}
where $m_{\rm gain}$ is the minimum held-out pooled improvement. The nominal and first recent-history partitions receive equal weight because they have nearly equal held-out counts and represent the frozen reference and causal short-term adaptation. The second recent-history partition is kept separate as a replication check rather than being averaged into the gain criterion. The executed macro action is
\begin{equation}
\hspace{-0.8em}
a_t^{\rm exec}
=
\begin{cases}
a_W,
& \hspace{-0.5em} \text{if all statistics are finite and \eqref{eq:rbpe_gate} holds},\\
\varnothing,
& \hspace{-0.5em} \text{otherwise},
\end{cases}
\label{eq:executed_macro}
\end{equation}
where $\varnothing$ invokes the selected causal fallback. The gate therefore uses data disjoint from candidate selection and requires support from two independently sampled recent-history partitions. It remains an empirical model-based admission rule, not a formal safety guarantee.

\subsection{Online Execution and Computational Complexity}
\label{subsec:execution_complexity}

For compactness, let $\mathsf{Eval}_a(S_t,\omega)$, $\mathsf{Eval}_{\rm MR}(S_t,\omega)$, and $\mathsf{Eval}_{\rm DPP}(S_t,\omega)$ denote the target-cohort returns obtained by applying macro rule $a$, Max-Rate, and DPP, respectively, during the first $L$ slots of scenario $\omega$, followed by a common suffix policy. Let $\mathsf{Eval}_{\mathcal A}$ denote joint macro-rule evaluation and $\mathsf{Eval}_{\mathcal B}$ joint fallback evaluation. The operators $\operatorname{Sel}$, $\operatorname{Depth}$, $\operatorname{Base}$, $\operatorname{Lower}$, and $\operatorname{Gate}$ perform selection, nested selection-depth assessment, fallback selection, held-out lower-statistic computation, and split-gate admission. Planning is invoked only when the unresolved active-job set is nonempty. Once admitted, a macro rule re-ranks the currently feasible
jobs in each of the next $L$ real slots rather than reserving a specific job in advance. If no macro rule is admitted, the selected fallback
policy determines the slot-level radio actions. In either case, the fixed GPU executor continues to evolve according to realized events, and the
virtual deficits are updated only from realized system outcomes. The fingerprint assertion in Algorithm~\ref{alg:rbpe_sr} verifies that
counterfactual rollout evaluation does not modify the realized system state, causal history, or virtual deficits.

Let $A=|\mathcal A|$ denote the number of candidate macro rules, $N_{\rm sel}$ and $N_{\rm gate}$ the numbers of selection and held-out futures, $H_t$ the maximum rollout horizon required to resolve the target cohort, $J_t$ the maximum number of active jobs in any rollout branch, and $G_t$ the per-slot cost of evaluating feasible GPU waves. Each selection future evaluates all $A$ macro rules and two fallback branches, whereas each held-out future evaluates only the frozen candidate and fallback. The serial planning complexity is therefore
\begin{equation}
\mathcal{O}\!\left(
\bigl((A+2)N_{\rm sel}+2N_{\rm gate}\bigr)
H_t(J_t+G_t)\right).
\label{eq:planning_complexity}
\end{equation}
When rollout branches are processed sequentially, the working-memory complexity is
\begin{equation}
\mathcal{O}(J_t+AN_{\rm sel}+N_{\rm gate}),
\label{eq:planning_memory}
\end{equation}
excluding immutable job traces, channel traces, and hardware profiles. Scenario and policy branches are mutually independent conditional on the frozen causal state and can therefore be evaluated in parallel, although parallelization does not reduce the total computational work.

\begin{algorithm}[t]
\caption{SGPE-SR}
\label{alg:rbpe_sr}
\begin{algorithmic}[1]
\Require $S_t,\mathcal F_t,\boldsymbol Z(t),
\mathcal A,\mathcal R,N_0^{\rm sel},N_1^{\rm sel},N^{\rm gate},L$
\Ensure $\{a^{\rm R}_{t+\ell}\}_{\ell=0}^{L-1}$

\State $\mathcal H_t\gets\operatorname{Sync}(\mathcal F_t)$
\State $X_t\gets
\operatorname{Fingerprint}(S_t,\mathcal H_t,\boldsymbol Z(t))$
\State $a_t^{\rm exec}\gets\varnothing$;\quad $B_t\gets{\rm MR}$

\If{$\mathcal J_t^{\rm act}\ne\varnothing$}
    \State $\Omega_t^{\rm sel}\perp\Omega_t^{\rm gate}
    \sim\{\widehat{\mathbb P}_r(\cdot\mid\mathcal F_t)\}_{r\in\mathcal R}$

    \ForAll{$\omega_{r,n}\in\Omega_t^{\rm sel}$}
        \State $\{R_{a,r,n}\}_{a\in\mathcal A}
        \gets\mathsf{Eval}_{\mathcal A}(S_t,\omega_{r,n})$
        \State $(R_{{\rm MR},r,n},R_{{\rm DPP},r,n})
        \gets\mathsf{Eval}_{\mathcal B}(S_t,\omega_{r,n})$
    \EndFor

    \ForAll{$N\in\{N_0^{\rm sel},N_1^{\rm sel}\}$}
        \State $a_N\gets\operatorname{Sel}
        \bigl(\{R_{a,{\rm N},n}:a\in\mathcal A,\,
        n\in\mathcal I_{\rm N}^{\rm sel}(N)\}\bigr)$
        \State $B_N\gets\operatorname{Base}
        \bigl(\{R_{{\rm DPP},r,n}-R_{{\rm MR},r,n}:
        r\in\mathcal R,\,1\le n\le N\}\bigr)$
    \EndFor

    \State $N_{\rm sel}\gets\operatorname{Depth}
    (a_{N_0},a_{N_1},B_{N_0},B_{N_1})$
    \State $(a_W,B_t)\gets(a_{N_{\rm sel}},B_{N_{\rm sel}})$

    \ForAll{$\omega_{r,n}\in\Omega_t^{\rm gate}$}
        \State $R^{\rm gate}_{a_W,r,n}
        \gets\mathsf{Eval}_{a_W}(S_t,\omega_{r,n})$
        \State $R^{\rm gate}_{B_t,r,n}
        \gets\mathsf{Eval}_{B_t}(S_t,\omega_{r,n})$
    \EndFor
    \ForAll{$r\in\mathcal R$}
        \State $L_r^{\rm gate}\gets\operatorname{Lower}
        \bigl(\{R^{\rm gate}_{a_W,r,n}-R^{\rm gate}_{B_t,r,n}:
        1\le n\le N_r^{\rm gate}\}\bigr)$
    \EndFor

    \If{$\operatorname{Gate}(L_{\rm N}^{\rm gate},
    L_{\rm R_1}^{\rm gate},L_{\rm R_2}^{\rm gate})=1$}
        \State $a_t^{\rm exec}\gets a_W$
    \EndIf
\EndIf

\State \textbf{assert}\;
$\operatorname{Fingerprint}(S_t,\mathcal H_t,\boldsymbol Z(t))=X_t$

\For{$\ell=0,\ldots,L-1$}
    \If{$a_t^{\rm exec}=\varnothing$}
        \State $a^{\rm R}_{t+\ell}\gets
        \pi_{B_t}(S_{t+\ell},\boldsymbol Z(t+\ell))$
    \Else
        \State $a^{\rm R}_{t+\ell}\gets
        \pi_{a_t^{\rm exec}}(S_{t+\ell})$
    \EndIf
    \State $S_{t+\ell+1}\gets
    F(S_{t+\ell},a^{\rm R}_{t+\ell},
    \xi^{\rm real}_{t+\ell};\pi_{\rm G})$
    \State $\boldsymbol Z(t+\ell+1)\gets
    \operatorname{Update}
    (\boldsymbol Z(t+\ell),\xi^{\rm real}_{t+\ell})$
\EndFor
\end{algorithmic}
\end{algorithm}

\section{Performance Evaluation}
\label{sec:evaluation}

This section evaluates SGPE-SR through an independently frozen long-horizon confirmation, load sensitivity, GPU-side mechanism diagnostics, and a separate method-design audit.

\subsection{Experimental Setup}
\label{subsec:simulation_setup}

A deterministic slot-level simulator reproduces the complete BU--SG--RU--RF--RH pipeline at 1-ms slot duration. Each confirmatory run contains 100~ms of warm-up and a 500-ms measurement interval; measurement jobs are followed until timely completion or deadline expiration. The warm-up exceeds the largest 75-ms deadline, and the measurement interval is five times longer than that used in the earlier revision analysis. All policies in a paired
seed receive identical workload and channel realizations. The main setting contains 16 tenants and two equiprobable GPU-compatibility labels. Different
labels cannot be co-batched, but share the same execution profiles. Jobs arrive as a Poisson process at an aggregate rate of 220 jobs/s. Four
four-tenant groups have reference rates of 75, 105, 140, and 185~Mbit/s, while the deadline classes are 45, 55, 65, and 75~ms. For tenant
$u\in\{1,\ldots,16\}$, the rate-group and deadline indices are $\lfloor(u-1)/4\rfloor$ and $(u-1)\bmod4$, so every rate group contains one
tenant from each deadline class. Per-tenant log-rate fading follows an AR(1) process with correlation 0.85 and innovation standard deviation 0.13, with
rates clipped to the supported range. To model input heterogeneity, $c_j$ is drawn from $\operatorname{Lognormal}(0,0.38^2)$, with the BU and RU payloads set to $150c_j^{0.25}$ and $525c_j$ kbit, respectively. The two lognormal parameters are the natural-log-space mean and variance. The reference case $c_j=1$ yields payloads of 150 and 525 kbit. Communication and computation scaling factors are one. GPU times are measured on an NVIDIA RTX 2000 Ada Generation Laptop GPU. The scout uses MobileNetV3-Small at $160\times160$, and refinement uses ResNet-18 split into front and head stages. FP16 isolated and two-stream concurrent execution is profiled at batch sizes 1, 2, and 4. Every radio policy uses the same P95 profiles and the same feasibility-aware, two-stream, non-preemptive GPU executor. The study is therefore a system-level Monte Carlo evaluation with measured GPU timing, not a live-system experiment. Development seeds 14901--14905 screened the operating load, and seeds 14611--14615 selected the gate margin before the long-horizon studies. A separately frozen design audit on previously unused seeds 200001--200030 compared the earlier distinct stress criterion with a recent-history-only alternative. Because the stress criterion did not improve the prespecified lower-tail metrics, it was removed and the method was renamed Split-Gate SGPE-SR before any final-confirmation outcome was generated. The final protocol, parameters, and seeds 200031--200060 were then frozen; all 30 previously unused confirmation seeds were retained. Secondary-load seeds 200061--200100 were assigned by load before the sensitivity runs. The main settings are summarized in Table~\ref{tab:table1}.

\begin{table}[t]
\centering
\footnotesize
\caption{Main System and SGPE-SR Parameters}
\label{tab:table1}
\begin{tabular}{lc}
\toprule
Parameter & Value\\
\midrule
Slot / warm-up / measurement & $1/100/500$ ms\\
Tenants / compatibility labels & $16/2$\\
Aggregate offered load & $220$ jobs/s\\
Reference rates & $75/105/140/185$ Mbit/s\\
Deadline classes & $45/55/65/75$ ms\\
Reference base / ROI payload & $150/525$ kbit\\
Complexity distribution & $\operatorname{Lognormal}(0,0.38^2)$\\
GPU statistic / maximum batch & P95 / 4\\
GPU execution streams & 2\\
AR(1) $\rho_{\rm ch}/\sigma_{\rm ch}$ & $0.85/0.13$\\
Context / arrival window / macro & $16/15/5$ slots\\
Total selection prefix / full / held-out gate &$32/64/64$\\
$\rho/\beta$ for DPP & $0.70/1$\\
$\delta/z_{\rm one}/m_{\rm gain}$ & $0.03125/1.64/0.046875$\\
Future-model partitions & $\mathrm N/\mathrm R_1/\mathrm R_2$\\
\bottomrule
\end{tabular}
\end{table}

\subsection{Baselines, Metrics, and Statistical Protocol}
\label{subsec:baselines_statistics}

Five causal baselines are considered. Feasibility-filtered Max-Rate serves the feasible job with the highest observed rate; feasibility-filtered earliest-deadline-first (EDF) uses the smallest
absolute deadline; and smallest-remaining-upload-first (SRUF) uses the smallest current payload. DPP follows \eqref{eq:deficit_update}--\eqref{eq:dpp_score}. GPU-aware scheduling combines deadline urgency, one-slot transmission progress, compatible ready-job count,
and remaining GPU stages. The optimistic feasibility filter is a common system-level eligibility mechanism, not part of any baseline's priority score. All policies share the filter, deterministic tie breaking, GPU executor, and exogenous stream instance.
For seed $i$ and policy $\pi$, let $Y_i^\pi$ be the number of measured jobs whose RH stage finishes by its deadline. For baseline $b$, the paired effect
is
\begin{equation}
\Delta_i^{(b)}=Y_i^\text{SGPE-SR}-Y_i^{(b)}.
\label{eq:paired_seed_effect}
\end{equation}
The primary comparison is SGPE-SR versus Max-Rate. A two-sided 95\% paired cluster-bootstrap percentile interval is computed from 100,000 resamples of
the 30 complete seeds. A two-sided exact sign-flip test is applied to the nonzero paired differences, while zero differences are retained as ties.
The primary bootstrap estimator is the unweighted mean of the 30 seed-level paired count differences in \eqref{eq:paired_seed_effect}. Aggregate completion ratios are descriptive ratios of summed timely jobs to summed offered jobs, not means of per-seed ratios. Additionally, relative timely-job gain, deadline-miss reduction, and win/loss/tie counts are reported. For a fixed offered workload, deadline misses equal offered jobs minus timely completions; timely-job gain and miss reduction are therefore two normalizations of one endpoint, not independent outcomes.

\subsection{Main System-Level Performance}
\label{subsec:main_results}

\begin{figure}[t]
\centering
\includegraphics[width=0.48\textwidth]{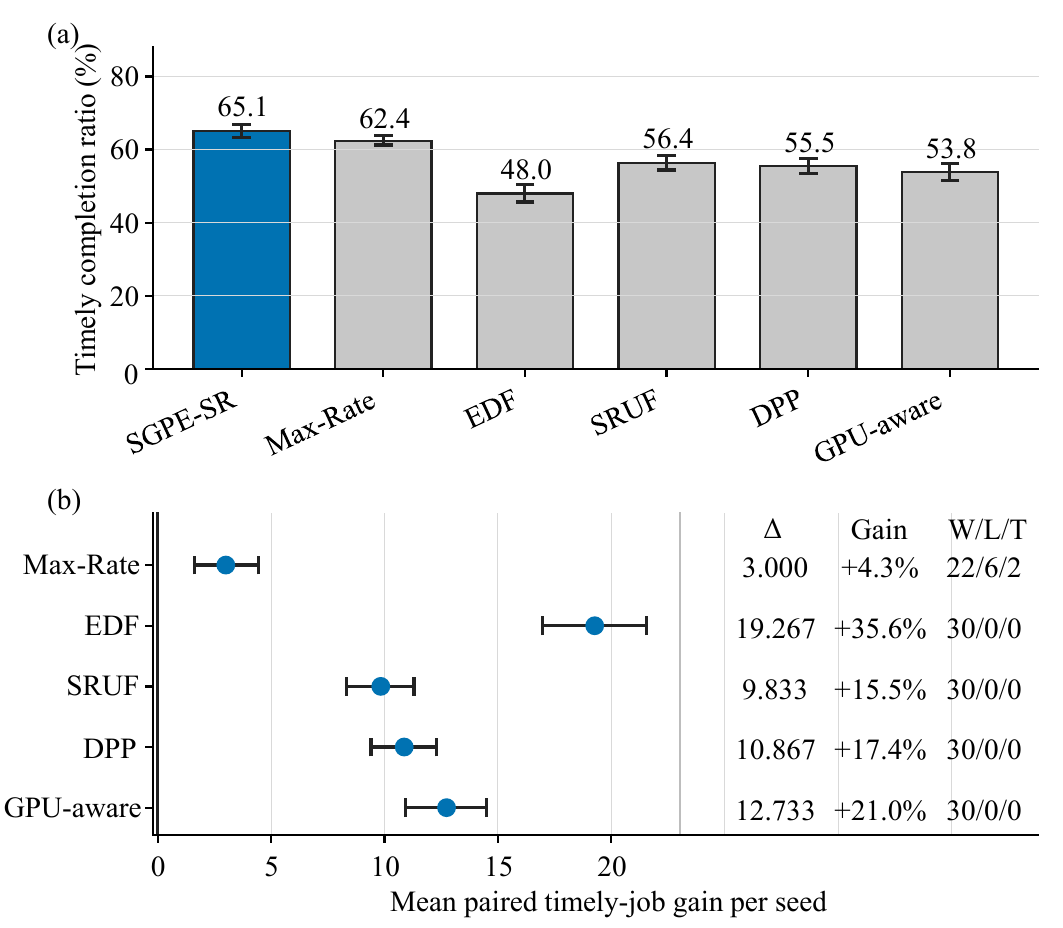}
\caption{Frozen long-horizon evaluation over 30 previously unused paired seeds: (a) aggregate timely-completion ratios; (b) mean paired timely-job gains of SGPE-SR over the baselines, with paired-bootstrap 95\% confidence intervals.}
\label{fig:main_system_performance}
\end{figure}

\begin{figure}[t]
\centering
\includegraphics[width=0.48\textwidth]{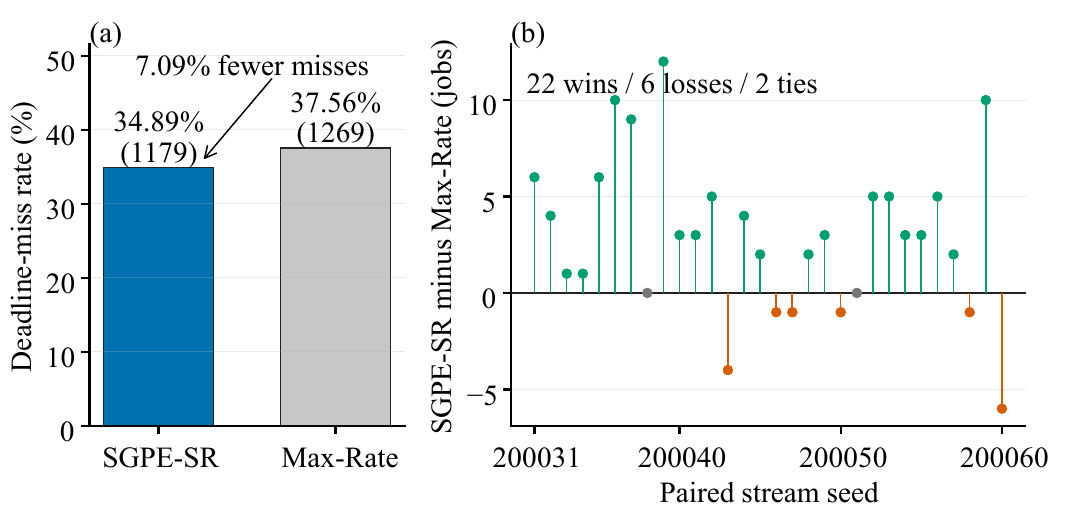}
\caption{Deadline outcomes relative to Max-Rate over the same 30 long-horizon confirmation seeds: (a) aggregate miss rates; (b) all seed-level differences in timely completions.}
\label{fig:deadline_reliability}
\end{figure}

Fig.~\ref{fig:main_system_performance} summarizes the final frozen evaluation across 30 previously unused seeds and 3,379 offered jobs. The controller, protocol, analysis rules, and seed set were fixed before any outcome from these seeds was generated. SGPE-SR completes 2200 jobs on time versus 2110 for Max-Rate, raising the aggregate completion ratio from 62.44\% to 65.11\% ($+2.66$ percentage points). This corresponds to 4.27\% more timely jobs and a mean paired gain of 3.00 jobs/run; the paired-bootstrap 95\% interval is $[1.60,4.43]$, and the exact sign-flip test gives $p=3.50\times10^{-4}$. The 22/6/2 win/loss/tie record shows that the effect is distributed across the frozen seeds. Relative timely-job gains over EDF, SRUF, DPP, and GPU-aware scheduling are 35.64\%, 15.49\%, 17.40\%, and 21.01\%, respectively. Max-Rate remains the strongest baseline: high-rate service shortens the immediate radio stage, but it does not account for whether an upload releases deadline-critical GPU work or changes subsequent non-preemptive waiting.

Fig.~\ref{fig:deadline_reliability} presents the same endpoint as deadline misses. SGPE-SR reduces misses from 1269 to 1179, lowering the aggregate miss rate from 37.56\% to 34.89\%, a 7.09\% relative reduction. Improvements occur in 22 seeds, ties in two, and degradation in six; the seed-level differences range from $-6$ to $+12$ timely jobs. Timely completions and misses are complementary counts, so this panel is a reliability-oriented visualization rather than an independent statistical outcome.
 
\subsection{Load Sensitivity}
\label{subsec:load_sensitivity}

\begin{figure}[t]
\centering
\includegraphics[width=0.48\textwidth]{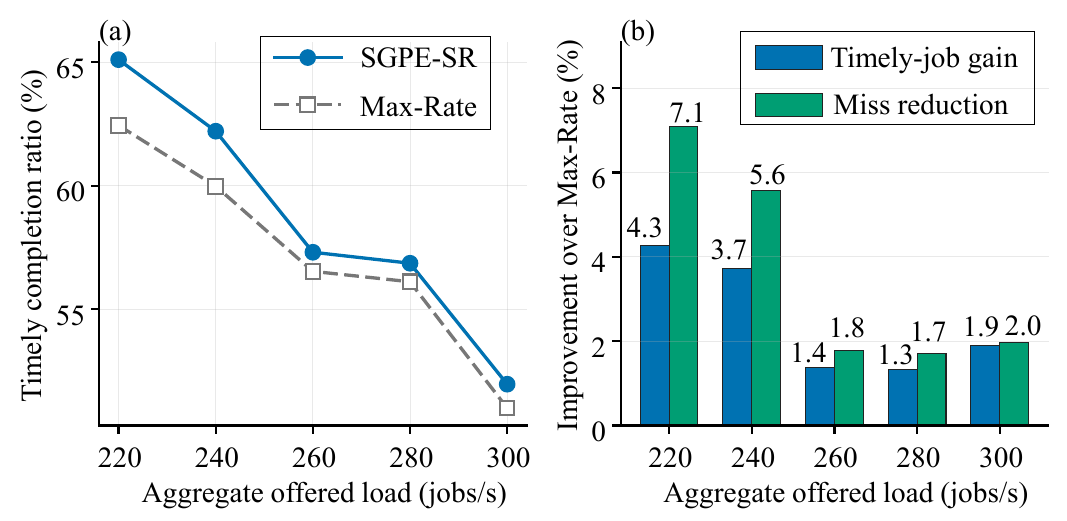}
\caption{Long-horizon sensitivity to aggregate offered load: (a) aggregate timely-completion ratios; (b) relative timely-job gain and deadline-miss reduction versus Max-Rate. The 220-jobs/s primary point uses 30 frozen confirmation seeds; each secondary point uses 10 seeds assigned before execution.}
\label{fig:load_sensitivity}
\end{figure}

Fig.~\ref{fig:load_sensitivity} extends the 100/500-ms long-horizon protocol to 220, 240, 260, 280, and 300 jobs/s. Aggregate timely-job gains over Max-Rate are 4.27\%, 3.71\%, 1.37\%, 1.34\%, and 1.89\%, respectively. At 240 jobs/s the 10-seed mean paired difference is 2.60 jobs/run with a 95\% interval of $[1.00,4.60]$; the intervals at 260, 280, and 300 jobs/s include zero. Thus, the point estimate is positive at every tested load, but the secondary points are sensitivity evidence with smaller samples and do not establish a significant benefit at every load.

\subsection{Coupling Mechanism and Method-Design Audit}
\label{subsec:mechanism_ablation}

\begin{figure}[t]
\centering
\includegraphics[width=0.48\textwidth]{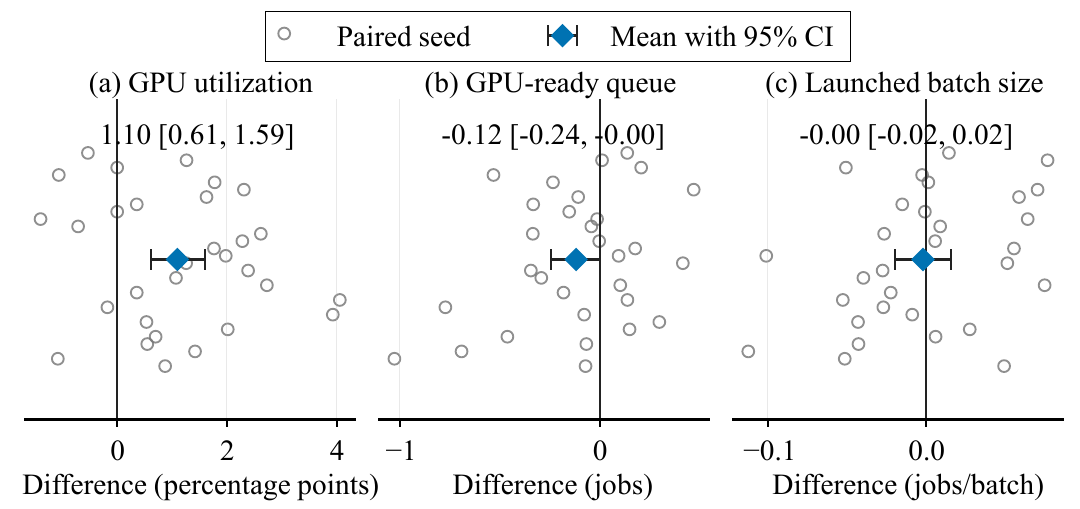}
\caption{GPU-side mechanism diagnostics relative to Max-Rate: (a) utilization, (b) mean ready-queue length, and (c) launched batch size. Circles denote paired-seed differences; diamonds and error bars denote means and paired-bootstrap 95\% confidence intervals.}
\label{fig:gpu_mechanism}
\end{figure}

Fig.~\ref{fig:gpu_mechanism} probes, rather than proves, the proposed radio--GPU coupling mechanism on the final confirmation seeds. Relative to Max-Rate, SGPE-SR increases mean GPU utilization by 1.10 percentage points, with a paired-bootstrap 95\% interval of $[0.61,1.59]$. The mean ready-queue length decreases by 0.118 job, with an interval of $[-0.243,-0.001]$, while the launched batch-size change is near zero ($-0.002$ job/batch) and its interval includes zero. The telemetry is consistent with earlier release of executable GPU work and shorter queues, but it does not isolate a unique causal mediator.

\begin{figure}[t]
\centering
\includegraphics[width=0.48\textwidth]{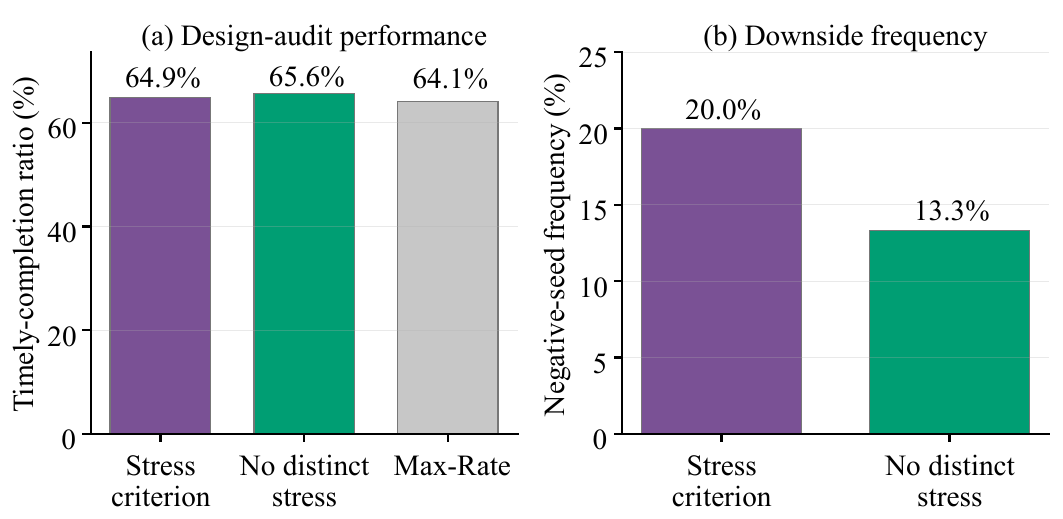}
\caption{Frozen long-horizon method-design audit over seeds 200001--200030: (a) aggregate timely-completion ratios for the earlier distinct-stress criterion, the alternative without a distinct stress model, and Max-Rate; (b) fractions of seeds with a negative timely-job difference versus Max-Rate.}
\label{fig:gate_ablation}
\end{figure}

Fig.~\ref{fig:gate_ablation} reports the design audit that preceded the final confirmation. The earlier distinct-stress controller achieves a 1.21\% gain over Max-Rate, with a paired interval of $[-0.40,2.03]$ jobs/run, whereas removing the distinct stress model yields a 2.38\% gain and an interval of $[0.50,2.77]$. The distinct-stress controller has negative seed-level differences in 20.0\% of seeds and a worst-20\% mean of $-4.50$ jobs/run; the corresponding values without the distinct stress model are 13.3\% and $-2.83$. Thus, the distinct stress criterion improves neither mean performance nor the prespecified lower-tail diagnostics. It was removed before seeds 200031--200060 were generated, and the final Split-Gate method was then evaluated once under the independently frozen protocol in Fig.~\ref{fig:main_system_performance}.

\section{Conclusion}
\label{sec:conclusion}

This paper investigated causal radio scheduling for deadline-constrained progressive edge inference, where each upload decision affects subsequent GPU-stage releases, batching opportunities, and non-preemptive waiting. The proposed SGPE-SR scheduler combines conditional stochastic rollouts, structured radio rules, a causal Max-Rate/DPP fallback selector, and independent selection and held-out gate samples. A frozen design audit found that a distinct stress-model criterion did not improve mean or lower-tail performance, so it was removed before the final confirmation. In the final frozen 30-seed evaluation with a 500-ms measurement interval, the resulting Split-Gate controller achieved 4.27\% more timely completions and 7.09\% fewer deadline misses than Max-Rate, with a positive paired confidence interval.Long-horizon sensitivity produced positive point estimates across 220--300 jobs/s, although the smaller secondary-load intervals did not all exclude zero.

\ifCLASSOPTIONcaptionsoff
  \newpage
\fi


\begin{thebibliography}{99}

\bibitem{Letaief2022EdgeAI}
K.~B. Letaief, Y. Shi, J. Lu, and J. Lu, ``Edge artificial intelligence for 6G: Vision, enabling technologies, and applications,''
\emph{IEEE J. Sel. Areas Commun.}, vol. 40, no. 1, pp. 5--36, Jan. 2022.

\bibitem{Shao2022TaskOriented}
J. Shao, Y. Mao, and J. Zhang, ``Learning task-oriented communication for edge inference: An information bottleneck approach,'' \emph{IEEE J. Sel. Areas Commun.}, vol. 40, no. 1, pp. 197--211, Jan. 2022.

\bibitem{Shao2023Cooperative}
J. Shao, Y. Mao, and J. Zhang, ``Task-oriented communication for multi-device cooperative edge inference,'' \emph{IEEE Trans. Wireless Commun.}, vol. 22, no. 1, pp. 73--87, Jan. 2023.

\bibitem{Chen2024Realtime}
Q. Chen, S. Guo, K. Wang, W. Xu, J. Li, Z. Cai, H. Gao, and A.~Y. Zomaya, ``Towards real-time inference offloading with distributed edge computing: The framework and algorithms,'' \emph{IEEE Trans. Mobile Comput.}, vol. 23, no. 7, pp. 7552--7571, Jul. 2024.

\bibitem{Shao2024VideoAnalytics}
J. Shao, X. Zhang, and J. Zhang, ``Task-oriented communication for edge video analytics,'' \emph{IEEE Trans. Wireless Commun.},
vol. 23, no. 5, pp. 4141--4154, May 2024.

\bibitem{Im2024AttentionAware}
J. Im, N. Kwon, T. Park, J. Woo, J. Lee, and Y. Kim, ``Attention-aware semantic communications for collaborative inference,''
\emph{IEEE Internet Things J.}, vol. 11, no. 22, pp. 37008--37020, Nov. 2024.

\bibitem{Li2023Adaptive}
P. Li, E. Koyuncu, and H. Seferoglu,
``Adaptive and resilient model-distributed inference in edge computing systems,''
\emph{IEEE Open J. Commun. Soc.},
vol. 4, pp. 1263--1273, 2023.

\bibitem{Lin2023CoInference}
X. Lin, R. Liu, J. Xie, Q. Wei, Z. Zhou, X. Chen, Z. Huang, and G. Lu, ``Online scheduling of CPU--NPU co-inference for edge AI tasks,''
in \emph{Proc. IEEE Wireless Commun. Netw. Conf. (WCNC)},Glasgow, U.K., Mar. 2023, pp. 1--6.

\bibitem{Lan2023Progressive}
Q. Lan, Q. Zeng, P. Popovski, D. G\"und\"uz, and K. Huang, ``Progressive feature transmission for split classification at the wireless edge,''
\emph{IEEE Trans. Wireless Commun.}, vol. 22, no. 6, pp. 3837--3852, Jun. 2023.

\bibitem{Li2024ModelSplitting}
X. Li and S. Bi, ``Optimal AI model splitting and resource allocation for device--edge co-inference in multi-user wireless sensing systems,''
\emph{IEEE Trans. Wireless Commun.}, vol. 23, pp. 11094--11108, 2024.

\bibitem{Lyu2024ObjectiveDriven}
X. Lyu, Y. Li, Y. He, \emph{et al.}, ``Objective-driven differentiable optimization of traffic prediction and resource allocation for split AI inference edge networks,'' \emph{IEEE Trans. Mach. Learn. Commun. Netw.}, vol. 2, pp. 1178--1192, 2024.

\bibitem{Liu2022VELTAIR}
Z. Liu, J. Leng, Z. Zhang, Q. Chen, C. Li, and M. Guo, ``VELTAIR: Towards high-performance multi-tenant deep learning services via adaptive compilation and scheduling,'' in \emph{Proc. 27th ACM Int. Conf. Architectural Support Program. Lang. Oper. Syst. (ASPLOS)}, 2022, pp. 388--401.

\bibitem{Cui2022DVABatch}
W. Cui, H. Zhao, Q. Chen, H. Wei, Z. Li, D. Zeng, C. Li, and M. Guo, ``DVABatch: Diversity-aware multi-entry multi-exit batching for efficient processing of DNN services on GPUs,'' in \emph{Proc. USENIX Annu. Tech. Conf. (USENIX ATC)}, 2022, pp. 183--198.

\bibitem{Choi2022Gpulet}
S. Choi, S. Lee, Y. Kim, J. Park, Y. Kwon, and J. Huh, ``Serving heterogeneous machine learning models on multi-GPU servers with spatio-temporal sharing,'' in \emph{Proc. USENIX Annu. Tech. Conf. (USENIX ATC)}, Carlsbad, CA, USA, Jul. 2022, pp. 199--216.

\bibitem{Han2022REEF}
M. Han, H. Zhang, R. Chen, and H. Chen, ``Microsecond-scale preemption for concurrent GPU-accelerated DNN inferences,'' in \emph{Proc. 16th USENIX Symp. Oper. Syst. Design Implement. (OSDI)}, Carlsbad, CA, USA, Jul. 2022, pp. 539--558.

\bibitem{Li2022Tetris}
J. Li, L. Zhao, Y. Yang, K. Zhan, and K. Li, ``Tetris: Memory-efficient serverless inference through tensor sharing,'' in \emph{Proc. USENIX Annu. Tech. Conf. (USENIX ATC)}, Carlsbad, CA, USA, Jul. 2022, pp. 473--488.


\bibitem{Ma2024ElasticRoom}
L. Ma, H. Chen, E. Shao, L. Wang, Q. Chen, and G. Tan, ``ElasticRoom: Multi-tenant DNN inference engine via co-design with resource-constrained compilation and strong priority scheduling,'' in \emph{Proc. 33rd Int. Symp. High-Perform. Parallel Distrib. Comput. (HPDC)},
2024, pp. 1--14.


\bibitem{Zhang2023Shepherd}
H. Zhang, Y. Tang, A. Khandelwal, and I. Stoica, ``SHEPHERD: Serving DNNs in the wild,'' in \emph{Proc. 20th USENIX Symp. Networked Syst. Design Implement. (NSDI)}, Boston, MA, USA, Apr. 2023, pp. 787--808.

\bibitem{Li2023AlpaServe}
Z. Li, L. Zheng, Y. Zhong, V. Liu, Y. Sheng, X. Jin, Y. Huang, Z. Chen, H. Zhang, J.~E. Gonzalez, and I. Stoica, ``AlpaServe: Statistical multiplexing with model parallelism for deep learning serving,'' in \emph{Proc. 17th USENIX Symp. Oper. Syst. Design Implement. (OSDI)}, Boston, MA, USA, Jul. 2023, pp. 663--679.

\bibitem{Kwon2023vLLM}
W. Kwon, Z. Li, S. Zhuang, Y. Sheng, L. Zheng, C.~H. Yu, J.~E. Gonzalez, H. Zhang, and I. Stoica,
``Efficient memory management for large language model serving with PagedAttention,'' in \emph{Proc. 29th ACM Symp. Oper. Syst. Princ. (SOSP)},
Koblenz, Germany, Oct. 2023, pp. 611--626.

\bibitem{Patel2024Splitwise}
P. Patel, E. Choukse, C. Zhang, A. Shah, I. Goiri, S. Maleki, and R. Bianchini, ``Splitwise: Efficient generative LLM inference using phase splitting,'' in \emph{Proc. 51st ACM/IEEE Annu. Int. Symp. Comput. Archit. (ISCA)}, Buenos Aires, Argentina, Jun. 2024, pp. 118--132.

\bibitem{Zhong2024DistServe}
Y. Zhong, S. Liu, J. Chen, J. Hu, Y. Zhu, X. Liu, X. Jin, and H. Zhang, ``DistServe: Disaggregating prefill and decoding for goodput-optimized large language model serving,'' in \emph{Proc. 18th USENIX Symp. Oper. Syst. Design Implement. (OSDI)}, Santa Clara, CA, USA, Jul. 2024, pp. 193--210.

\bibitem{Agrawal2024Sarathi}
A. Agrawal, N. Kedia, A. Panwar, J. Mohan, N. Kwatra, B. Gulavani, A. Tumanov, and R. Ramjee,
``Taming throughput--latency tradeoff in LLM inference with Sarathi-Serve,'' in \emph{Proc. 18th USENIX Symp. Oper. Syst. Design Implement. (OSDI)},
Santa Clara, CA, USA, Jul. 2024, pp. 117--134.

\bibitem{Sun2024Llumnix}
B. Sun, Z. Huang, H. Zhao, W. Xiao, X. Zhang, Y. Li, and W. Lin, ``Llumnix: Dynamic scheduling for large language model serving,''
in \emph{Proc. 18th USENIX Symp. Oper. Syst. Design Implement. (OSDI)}, Santa Clara, CA, USA, Jul. 2024, pp. 173--191.


\bibitem{Ren2024PredictiveUplink}
C. Ren and X. Lyu,
``Online-learning-based predictive optimization of uplink scheduling for industrial Internet-of-Things,'' \emph{IEEE Open J. Commun. Soc.},
vol. 5, pp. 6817--6831, 2024.

\bibitem{Sfaxi2024ProactivePlacement}
H. Sfaxi, I. Lahyani, S. Yangui, and M. Torjmen, ``Latency-aware and proactive service placement for edge computing,'' \emph{IEEE Trans. Netw. Serv. Manag.}, vol. 21, no. 4, pp. 4243--4254, Aug. 2024.

\bibitem{Wang2024DigitalTwin}
Y. Wang, J. Fang, Y. Cheng, H. She, Y. Guo, and G. Zheng, ``Cooperative end--edge--cloud computing and resource allocation for digital twin enabled 6G industrial IoT,'' \emph{IEEE J. Sel. Topics Signal Process.}, vol. 18, no. 1, pp. 124--137, 2024.

\bibitem{Agrawal2024Vidur}
A. Agrawal, N. Kedia, J. Mohan, A. Panwar, N. Kwatra, B.~S. Gulavani, R. Ramjee, and A. Tumanov, ``VIDUR: A large-scale simulation framework for LLM inference,'' in \emph{Proc. Mach. Learn. Syst. (MLSys)}, vol. 6, pp. 351--366, 2024.




\end{thebibliography}
\end{document}